UNIVERSITÄT BERN

Faculty of Science
Institute of Computer Science
**Research Center for Digital Sustainability**

Bern, 30 June 2021

# Security implications of digitalization: The dangers of data colonialism and the way towards sustainable and sovereign management of environmental data

Report for the Federal Department of Foreign Affairs FDFA

**Authors:**
PD Dr. Matthias Stürmer, Dr. Jasmin Nussbaumer, Pascal Stöckli

Research Center for Digital Sustainability
Institute of Computer Science
University of Bern

www.digitale-nachhaltigkeit.unibe.ch

# Table of Content

# Glossary

**Big tech:** companies like Google, Apple, Microsoft, Amazon and Facebook (GAFAM)

**Data colonialism/digital colonialism:** control of data and digital infrastructure by big tech corporations

**Data repository:** traditional relational databases as well as new forms of data storage facilities

**Data sovereignty:** data is subject to the laws and governance structures within the nation where it is collected

**Digital skills:** programming and data science skills that allows people to use and create digital tools

**Digital self-determination:** individuals are able to control and use their data and any other personal digital assets in a self-determined way involving privacy, data protection and freedom of choice

**Digital sovereignty:** data, computing infrastructure, security, networks and any other digital topic is under control of the country or individual, not of big tech companies or other ICT corporations

**Digital sustainability:** unrestricted access to digital knowledge such as open data, crowd-sourced data and content, open source software, or open hardware for the entire society that provide positive ecological, social or economic effects

**Environmental data:** data that measures, models, or monitors various elements of the environment, especially satellite imagery and other geospatial data

**eu-LISA:** European Agency for the operational management of large-scale IT Systems in the area of freedom, security and justice

**Google Earth Engine:** cloud computing platform for processing satellite imagery and other geospatial data

**Hyperscale data center:** large data center run by a big tech corporation or another large ICT company

**PlanetLabs:** private American earth imaging company

**Swiss Data Cube:** non-for-profit IT platform of earth observation data (satellite images etc.) providing access to large spatio-temporal data in an analysis ready form for environmental monitoring and reporting

**Vendor lock-in:** dependence of customers on the products or services of a certain vendor

# Abbreviations

**API:** Application Programming Interface

**ESRI:** "Environmental Systems Research Institute", an American vendor of GIS software such as ArcGIS

**ICT:** Information and Communication Technology (including telecom sector)

**IT:** Information Technology (without telecom sector)

**GAFAM:** Google/Alphabet, Amazon, Facebook, Apple, and Microsoft

**GDP:** Gross Domestic Product

**GIS:** Geographical Information Systems

**GTC:** General Terms and Conditions

**OGD:** Open Government Data

**MEA:** Multi-lateral Environmental Agreement

**NGO:** Non-governmental Organization

**RBO:** River Basin Organization

**SEO:** Search Engine Optimization

**SDG:** Sustainable Development Goal

**SLA:** Service Level Agreement

**SOMO:** Stichting Onderzoek Multinationale Ondernemingen (Centre for Research on Multinational Corporations)

**UNEP:** United Nations Environment Program

**UNSCD:** United Nations Commission for Social Development

**WASH:** Water, Sanitation and Hygiene Sector

# Management summary

Digitalization opens up new opportunities in the collection, analysis, and presentation of data which can contribute to the achievement of the 2030 Agenda and its Sustainable Development Goals (SDGs). In particular, the access to and control of environmental and geospatial data is fundamental to identify and understand global issues and trends. Also immediate crises such as the COVID-19 pandemic demonstrate the importance of accurate health data such as infection statistics and the relevance of digital tools like video conferencing platforms. However, today much of the data is collected and processed by private actors. Thus, governments and researchers depend on data platforms and proprietary systems of big tech companies such as Google or Microsoft. The market capitalization of the seven largest US and Chinese big tech companies has grown to 8.7tn USD in recent years, about twice the size of Germany's gross domestic product (GDP). Therefore, their market power is enormous, allowing them to dictate many rules of the digital space and even interfere with legislations.

Based on a literature review and nine expert interviews this study presents a framework that identifies the risks and consequences along the workflow of collecting, processing, storing, using of data. It also includes solutions that governmental and multilateral actors can strive for to alleviate the risks. Fundamental to this framework is the novel concept of "data colonialism" which describes today's trend of private companies appropriating the digital sphere. Historically, colonial nations used to grab indigenous land and exploit the cheap labor of slave workers. In a similar way, today's big tech corporations use cheap data of their users to produce valuable services and thus create enormous market power.

The major dilemma is that many of the technically sophisticated services offered by big tech companies are very cheap or even seemingly free. Through their huge hyperscale data centers and with their highly skilled workforce, they are able to provide high-end information communication technology (ICT) infrastructure as well as advanced user-friendly computing services at a very low or no price. However, not only the big tech corporations, also smaller but focused software vendors create dependence of governmental actors and thus increase the growing asymmetry of knowledge and digital skills between the private and the public sector.

The interviews in this report show that Western countries and also developing countries today depend heavily on digital products and services from large and medium-sized IT enterprises, mostly located in the US and China. The powerful position of such companies leads to a weak negotiating position of public actors and thus often to an uncritical use of convenient digital services without further reflection on the long-term consequences. The situation resembles a seemingly relaxed existence inside a gilded cage controlled by powerful tech corporations.

This report summarizes historical as well as current academic and practitioner-oriented literature regarding the problem of data colonialism and the loss of digital sovereignty. The analysis of nine expert interviews from Swiss government, UN organizations, NGOs and academics draws a comprehensive picture of key issues facing society in the digital space today and in future. Besides elaborating the problem of IT provider dependence and illustrating current cases of data colonialism, this report also highlights solutions regarding data sovereignty and draws a path towards digital sustainability of the future virtual space.

Several examples in the interviews show how governmental actors and academics are able to regain control of their data and infrastructure when they invest into government controlled data analytics platforms, and build up common standards, data repositories, software and computing sites. One key element is supporting and relying on collaborative digital platforms such as the openly licensed, crowd-sourced global geographical information data community OpenStreetMap or the governmental and scientific driven environmental monitoring infrastructure Swiss Data Cube.

Therefore, the authors of this report recommend that professional IT users like governmental bodies and researchers become aware of the long-term problems of relying on data analytics, software development and IT infrastructure provided by private corporations. Consequently, the solutions include recommendations such as using, investing into and releasing open source software and publicly owned IT infrastructure. Also, governments should increase their employer appeal to attract young data science and software developer talent.

# 1 Introduction

Countries from all over the world have benefit from digitalization as illustrated by statements at the 59th session of the United Nations Commission for Social Development (UNCSD) in February 2021 named "A socially just transition towards sustainable development: The role of digital technologies on social development and well-being of all" (UNCSD, 2021). The COVID-19 pandemic has further emphasized the need for people's access to digital technologies and accurate, up-to-date data for decision makers. For example, in Guyana, the government has offered virtual help desks to report domestic violence and child abuse while people were not allowed to hold physical meetings during the pandemic. Or in Ethiopia, digital technology has improved effective interaction among citizens, government and businesses through a new Ethio-Migrant Database. Many of the speakers pointed out the challenge of the growing digital divide producing a new class of 'digital poor'. Therefore, in Malaysia the educational sector has promoted digital literacy leading to more than 3.1 million students and 387'000 teachers now using Google Workspace for Education.

These examples of technology use during the pandemic illustrate a growing trend that started long before the COVID-19 outbreak: Often, governments are confronted with immediate challenges in the digital space and quickly find convenient solutions by using IT services and infrastructure of big tech corporations and also smaller IT companies. As a result of this, the economic influence of tech companies has grown significantly in recent years. The new report "Engineering digital monopolies: The financialization of Big Tech" by the Centre for Research on Multinational Corporations (Stichting Onderzoek Multinationale Ondernemingen, SOMO) in December 2020 illustrates the enormous market power of the mostly US and Chinese owned big tech companies (Fernandez et al., 2020). These companies dominate the global digital sphere, some of them valuing more than the GDP of most countries of this world.

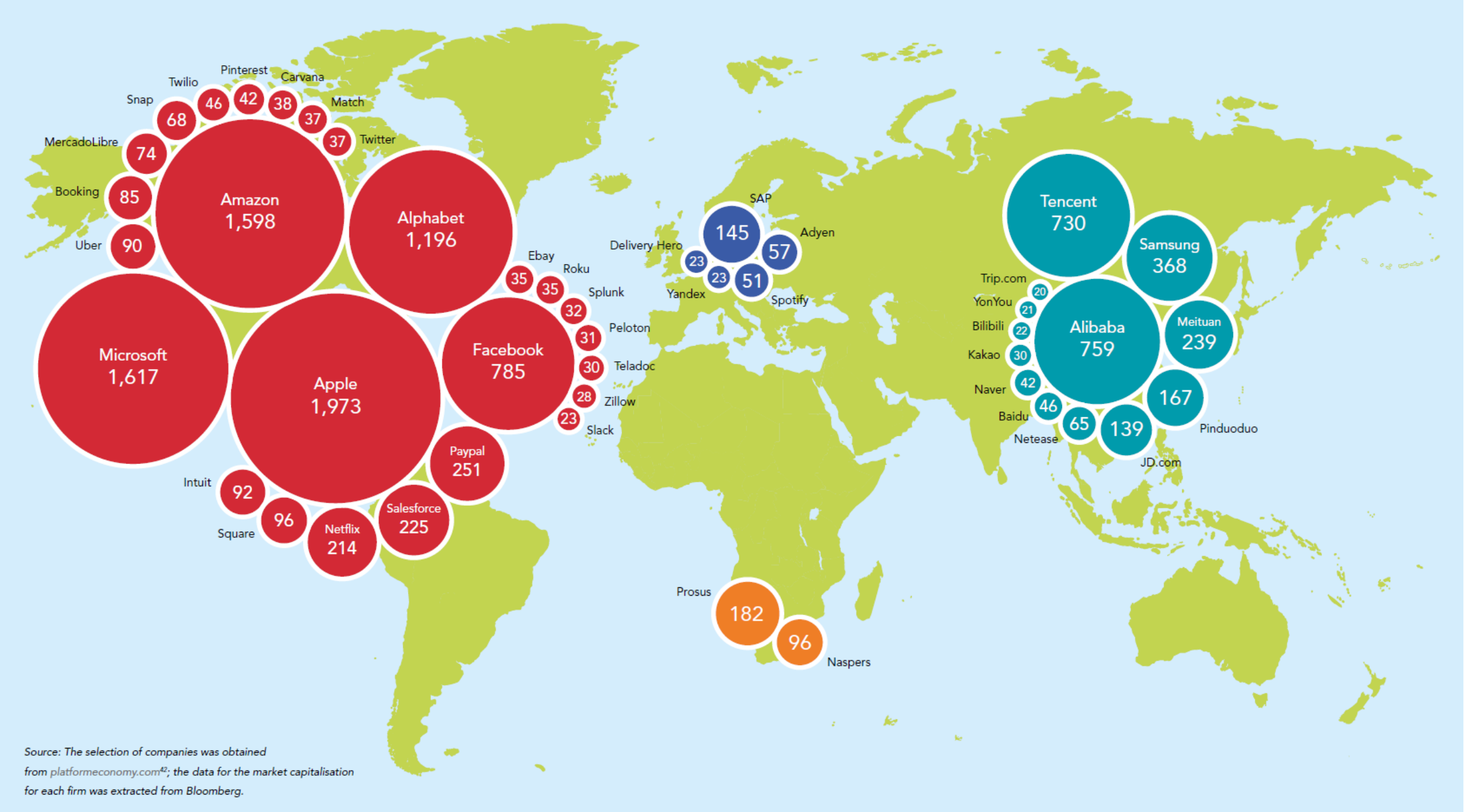

*Figure 1: Big Tech market capitalization (above US$ 20 billion) in December 2020*
*(Source: SOMO Report 2020 by Fernandez et al.)*

The Swiss government recently issued the first Digital Foreign Policy Strategy addressing digital governance, sustainable development, Cybersecurity, and digital self-determination (Federal Department of Foreign Affairs, 2020). In particular, the aim of strengthening digital self-determination connects well with this report which expands on the topics data control and digital sovereignty.

The report continues with a literature overview introducing historical publications on Internet governance and explaining the concepts of data colonialism, data sovereignty, digital sovereignty, and digital sustainability particularly in relation to environmental data. Subsequently, an overview of all interviewed academics and governmental practitioners is presented (the summaries of the nine expert interviews can be found in Appendix B). In the analysis section a framework of data workflow, risks, consequences and recommendations is elaborated. Finally, a conclusion section summarizes the findings and provides an outlook with further topics to investigate.

# 2 Literature overview

Control of governmental digital activities by private companies is a long-standing issue. Since the rise of the Internet in the late 1990s IT corporations govern the cyberspace by means of their technological leadership as Harvard Law School professor Lawrence Lessig describes in his books "Code" (Lessig, 2000, 2006). Being a law professor, he points out that in cyberspace "code is law", meaning that regulation of the Internet happens through the source code of software developers. In another seminal book, "The Wealth of Networks", Lessig's colleague Yochai Benkler expands his concept of "commons-based peer production" towards the cyberspace to explain collaborative efforts of sharing information and creating new digital artifacts (Benkler, 2008). His notion of a "networked information economy" is based on the work by Nobel Prize winner Elinor Ostrom on common pool resources (Ostrom, 1990). Such common pool resources are not controlled by the government or the free market but by members of the civil society.

## 2.1 The power of Google, Apple, and Facebook

Nevertheless, the private sector has grown tremendously during the last ten years and now controls much of the Internet's data and IT infrastructure (Fernandez et al., 2020). Recently scholars started using the term "digital governmentality" to emphasize the logics of power and control on the Internet (Badouard et al., 2016; Barry, 2019). Similar to the early days of the cyberspace when private corporations defined how governments must behave on the web, they now extend their rules to Internet search and mobile device management. For instance, Google directs how websites have to be designed in order to optimize the indexing with their search engine. There is an entire stream of scientific research regarding Search Engine Optimization (SEO) that focuses on the unregulated corporate decisions of Google on how to rank content on the Internet (Ziakis et al., 2019).

Another example concerns the way of how to install applications on mobile devices. Apple and Google control the app environment of the mobile operating systems iOS and Android, thus defining the rules of the game in the app stores. This is being exploited e.g. in particular by Apple with its App Store because that is the only access for iPhone and iPad users to install software (Shoemaker, 2019). This confinement of customers has prompted the European Union to issue a formal antitrust case against Apple in 2020 following complaints by Spotify that Apple abuses its market power (Geradin and Katsifis, 2020).

Tech corporations have also played an important role within the public health care sector (Storeng and Puyvallée, 2021). For example by using their smartphones people could participate in digital contact tracing during the COVID-19 pandemic. In a historic joint intervention Google and Apple launched the "Google-Apple exposure notification system" (GAEN) in April 2020, soon after the outbreak of the pandemic. Inspired by a technical concept called "Decentralized Privacy-Preserving Proximity Tracing" developed by Swiss researchers (Troncoso et al., 2020) Google and Apple introduced a new interoperable application programming interface (API) to exchange anonymous proximity information between mobile devices. Although Google and Apple did not receive any direct benefit or data access, this example illustrates how dependent society has become on the big tech industry. The swift cooperation of the two corporations was in the public's interest, because otherwise the technical foundations for the COVID-19 tracing apps would not have worked.

Recent cases involving Facebook demonstrate again how a company is able to use its market power to enforce user behavior or even pressurize politics in the legislation process. For instance, in January 2021 Facebook announced that the new terms of service of WhatsApp would allow them to integrate user account registration data of WhatsApp users into Facebook accounts thus extending the ability of the company to construct user profiles for marketing purposes (Gomez, 2021). Or in February 2021 Facebook requested the Australian Government to change its media law since it would have forced the US company to pay for showing Australian news on its platform (Khalil, 2021). When the government did not do so, Facebook blocked Australian news from being shared on its platform, which put pressure on the country's legislation process (Porter, 2021). These examples illustrate how large tech companies control much of the digital space and show the drastic measures they are willing to take to enforce their commercial interests, when they appear at stake. This situation is the starting point for a comparison of activities by the tech industry with a historical epoch, the colonial period.

## 2.2 Concepts of data colonialism and digital sustainability

The concept of data colonialism has been elaborated by Couldry and Mejias (2019a, 2019b, 2019c) comparing the age of colonialization since the 15th century with today's technological conquests by big tech corporations. Similar to European countries colonizing land and people in Africa, the Americas and Asia, nowadays American and Chinese tech companies are conquering personal and governmental data of the global population. Couldry and Mejias draw direct parallels between the two eras regarding appropriation of resources (land/gold vs. data), big economic profit for the colonizing powers, and their 'positive' ideologies to cover up real problems.

During historical colonialism, the European powers acquired cheap territory and extracted natural resources such as gold from their colonies with slaves (cheap labor). Today, big tech corporations are grabbing cheap data from the people as raw material for their "cloud empires". Couldry and Mejias therefore conclude: "Like cheap nature, cheap social data can be fully capitalized only through the exploitation of cheap labor." These companies do this with considerable public relations and marketing for the good cause. For instance, Facebook claims "we connect people" and Google's mission is "to organize the world's information and make it universally accessible and useful." Similarly, the colonial nations aspired for ideologies that were seemingly desirable, such as the civilization of indigenous peoples.

Kwet (2019) uses the similar term 'digital colonialism' analogous to 'data colonialism' as conceptual framework to describe economic domination by US Big Tech corporations. The 'tech hegemony' by GAFAM (Google/Alphabet, Amazon, Facebook, Apple, and Microsoft) is growing based on the support of the US government and thus the United States reinvents colonialism in the Global South. Digital colonialism can be countered by 'digital sovereignty' i.e. restoring control and ownership of key information and communications infrastructures (Pinto, 2018). Digital sovereignty involves all digital infrastructures (IT hardware and network) while the much older notion of 'data sovereignty' mainly focuses on the geographical location of data in the cloud (Peterson et al., 2011).

In 2017, one of the authors of this report introduced the concept of 'digital sustainability' linking the long-term availability of data and software with the aim for sustainable development (Stuermer et al., 2017). The notion of digital sustainability integrates with the aim for 'data sovereignty' and 'digital sovereignty' that increases the control of the individuals for their data and other digital assets. In addition, digital sustainability explains how data digital artifacts (software and data) are being produced and used by stakeholders within an ecosystem. Software development in open source communities is an example of long-term, distributed production of digital knowledge commons. Digital sustainability therefore addresses the longevity of and sovereignty over digital knowledge and digital infrastructure, thus presenting another strategy to reduce the problems of data colonialism.

# 3 Interviews

In order to assess the various concepts in the literature above and link them to real-world examples, 9 expert interviews were conducted following a semi-structured guideline. This section provides an overview of the experts and the questions. The summaries of all interviews can be found in Appendix B.

The persons listed in Table 1 were interviewed for this study. They were selected according to their scientific background on environmental data or knowhow on digitalization issues. All of them belong to Swiss governmental or multilateral agencies, academic institutions, or non-governmental organizations.

| Date, duration | Name | Role and Organization | Special Interests |
|---|---|---|---|
| 29 January 2021, 1h 31min | **David Jensen** | Head of the UNEP Digital Transformation Task Force, Head of Policy and Innovation, Crisis Management Branch UN Environment | Business models for producing digital public goods, environmental data governance framework |
| 8 February 2021, 1h 7min | **Dr. Matthias Leese** | Senior researcher at the Center for Security Studies (CSS) at ETH Zürich | Predictive Policing |
| 11 February 2021, 39min | **Anna Brach** | Head of Human Security at Geneva Centre for Security Policy (GCSP) | Environmental and health security, global public commons and resource management |
| 17 February 2021, 58min | **Dr. Fritz Brugger** | Senior scientist at NADEL - Center for Development and Cooperation of ETH Zürich | Natural resource extraction, role of extractive companies |
| 18 February 2021, 1h 30min | **Prof. Stefan Keller** | Professor at Institute for Software at Ostschweizer Fachhochschule | Geo Information Systems (GIS), Data Engineering, Spatial Data Analytics, OpenStreetMap Expert |
| 19 February 2021, 1h 1min | **Thomas Schneider** | Ambassador and Director of International Relations, Swiss Federal Office of Communication (OFCOM), chairman of Swiss IGF Steering Group | Internet and digital governance |
| 22 February 2021, 1h 13min | **PD Dr. Andreas Heinimann** | Head of Regional Stewardship Hubs at the Wyss Academy for Nature; Associated Senior Research Scientist of the Centre for Development and Environment (CDE) of the University of Bern | Sustainable regional development, knowledge production, development interventions, and policy |
| 12 March 2021, 58min | **Prof. Nick Couldry and Prof. Ulises A. Mejias** | Professor of Media, Communications and Social Theory at London School of Economics and Political Science / Associate Professor of Communication Studies and Director of the Institute for Global Engagement at the State University of New York, College at Oswego. | Data colonialism, platform capitalism, authors of the book "The Costs of Connection – How Data Is Colonizing Human Life and Appropriating It for Capitalism" |
| 17 March 2021, 1h 11min | **Dr. Fatine Ezbakhe** | Scientific Officer at University of Geneva and the Geneva Water Hub | Environmental data governance, water data |

*Table 1: List of interviewees for this report*

The semi-structured interview guideline covered questions in three different areas (see Appendix A for full set of questions): First, there were questions about the individual perspective, asking what data the experts use for their own research activities, where they store it, and what experience they have with managing the data and digital platforms they use. The second block of questions addressed governmental data access and data use from a global perspective requesting the experts experience on business models of technology corporations or control of data by the private sector. Third, the questions focused on the concepts of data colonialism, digital sovereignty and digital sustainability asking the experts on their opinion about these terms and what potential solutions and exemplary platforms and initiatives they were aware of.

All interviews were conducted via video call and were recorded. The interviews were then transcribed verbatim and analyzed using a qualitative data analytics software. The interview summaries in Appendix B are written in rather narrative style to maintain the personal character of the conversations.

# 4 Analysis

The nine interviews illustrate today's importance of environmental data and the critical role of technology companies with various practical examples. They emphasize the widespread use of cloud-based solutions such as Google Earth Engine including the benefits and risks involved with such platforms. The interviews also show already existing initiatives as well as future solutions that would lead to more data sovereignty and digital sustainability. Linking these empirical insights with the literature above reveals several risks and patterns of data colonialism: For example, data and programming scripts are appropriated by the ICT industry resulting in vendor lock-in situations and decreasing data sovereignty of users.

Merging the referenced literature with the conducted interviews, a framework of data workflow, risks, consequences and recommendations is developed and described next (see Figure 2): The *data workflow* outline which steps take place in the process of data collection, data transformation and storage, data use, and the impact of data. The *risks* show different threats looming at the levels of data sources, data processing, and information aggregation. The *consequences* describe the effects evolving from the risks. And the *recommendations* propose solutions on how to address the risks and improve the situation in the public interest.

## 4.1 Data workflow

The data workflow is aligned with the data, information, and knowledge pyramid of information systems research (Rowley, 2007; Ackoff, 1989). In this notion, data is being processed into information that eventually leads to knowledge.

Therefore as a first step in the data workflow, raw data is being collected from various sources. For example environmental data includes satellite data provided by governmental organizations (e.g. NASA or ESA) and private companies (e.g. PlanetLabs). Additional data from measurements on the ground is collected by local or national governments, international organizations, NGOs, scientists or the civic society.

As second step, data is then processed into information either by commercial service providers or by non-commercial organizations and communities. Often data is being stored and processed on platforms of private companies since they offer flexible, user friendly and highly performing online services. Non-commercial providers regularly lack long-term funding, are less performing, or lower user experience than commercial platforms.

The results of data processing are environmental data sets containing information about earth surface, weather and climate, water, land use, mobility data etc. This information is then used as knowledge for emergency and disaster response, during humanitarian conflicts, for wildlife monitoring, climate change, predictive policing etc. Such environmental data impacts on forming new governmental policies, improves contract negotiations among countries and other parties, and leads to data-driven actions of various actors.

## 4.2 Risks

The following environmental, economic and social risks evolve from the data workflow illustrated above. These risks were mentioned several times in the expert interviews. They are relevant at three different levels and can therefore be divided into three groups of risks: Risks at the level of A) data sources, risks of B) data processing into information, and risks of C) information aggregation into knowledge.

**A) Risks at the level of data sources**

Data about the environment, the population, the economy etc. is collected by national and regional authorities as well as local communities, international and non-governmental organizations, researchers and scientific committees, and the civic society. In the environmental sector there are large governmental organizations such as NASA or ESA that collect geospatial data with their own satellite systems. In addition there are many private data collectors such as Google or PlanetLabs that create earth observation data and satellite images. Currently, no regulations regarding geospatial data collection exist, which means there is no limitation on what companies may or may not do with the data, whom they are allowed to sell it, and what strategic advantages they can draw out of it.

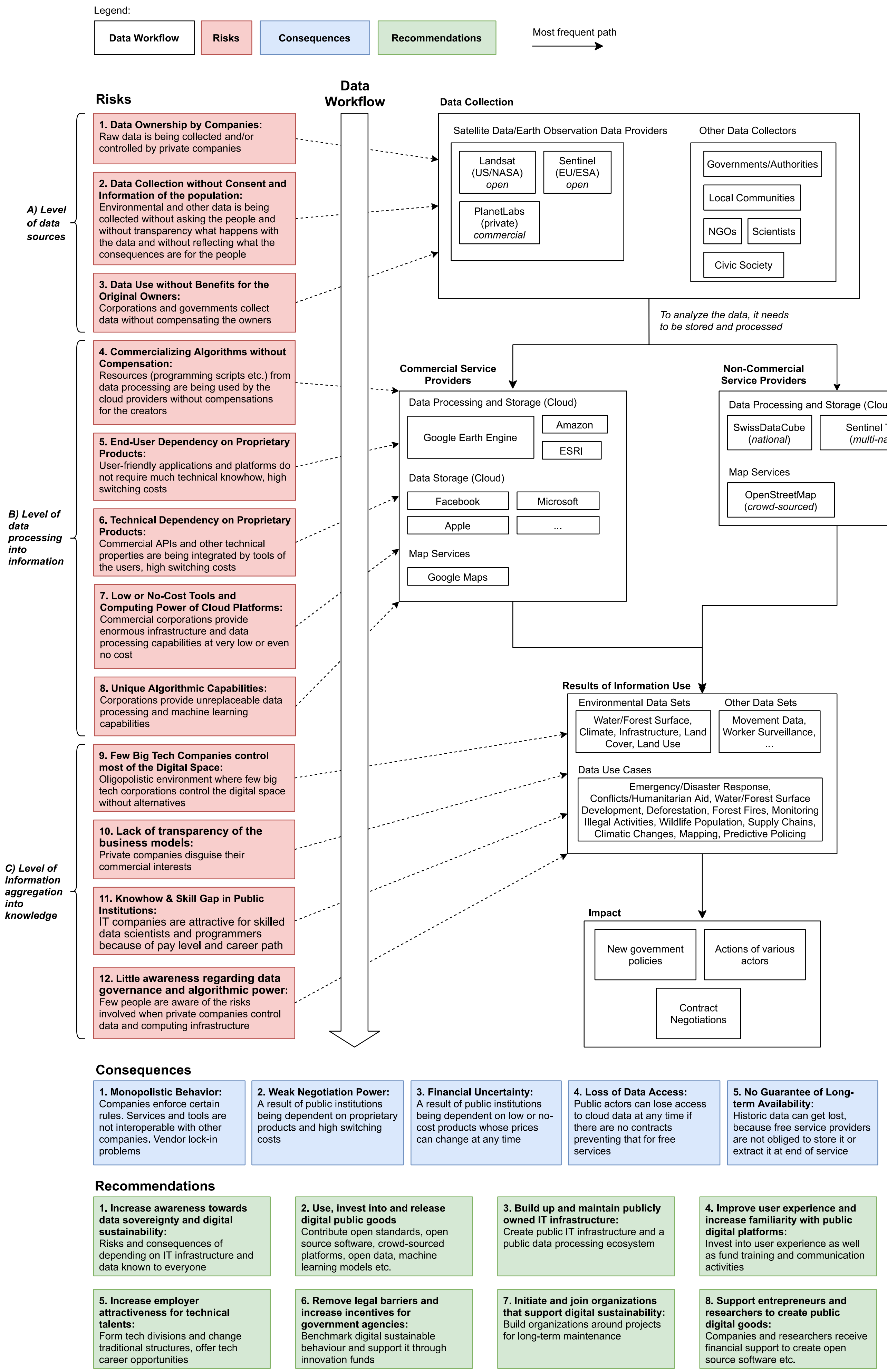

*Figure 2: Framework of data workflow, risks, consequences and recommendations*

**1. Data ownership by private companies**

If raw data is collected by private companies via user generated data or own technologies, the data providers are in full control of its processing and use. As data owners they determine if they allow others to access the data or not, and if yes, under what commercial conditions. By managing technical data access through online platforms or application programming interfaces (API) and by letting users sign general terms and conditions (GTC) they control in detail who can use the data for what purpose. As a bonus the private data owners receive valuable access statistics from their platform clients that reveal usage patterns leading to new insights and indicating potential business opportunities. If governments use such privately held data they are depend fully on the commercial goals and strategies of the corporations.

**2. Data collection without consent and information of the population**

Frequently, environmental data and other information is collected without consent of the people or the governments e.g. when conducting household surveys or mapping their land or habitats. Data gathering usually takes place without transparency of what happens with the data and without consideration of the consequences for the different stakeholders. On an aggregated level there is the ethical risk that data collectors use the data not in the interest of the inhabitants but for their commercial or political goals. This demonstrates the concept of data colonialism.

**3. Data use without benefits for the original owners**

Corporations and governments normally collect and use data without compensating the original owners of the data. This happens e.g. with local communities when their land is being mapped, with business owners who use mobile apps for credit loans, or with researchers who upload their programming scripts to the Google Earth Engine. Data appropriation thus takes place with minimal or without any benefit for the data owners. This again shows the characteristics of data colonialism where users provide cheap data as property of corporations for their commercial purposes.

**B) Risks at the level of data processing into information**

Raw data is being stored and processed on IT infrastructure in order to gain new information. Platforms like Google Earth Engine, cloud providers like Amazon Web Services, or geographical information systems (GIS) vendors like ESRI offer practical tools to transform and analyze data. They all offer online services that expect the users to upload their data on the servers of the companies called "the cloud".

**4. Commercializing algorithms without compensation**

Users of the online platforms upload data and programming scripts to cloud providers who are able to reuse these digital assets without compensating their original creators. Thus they basically hand over their intellectual property to private companies who are able to commercialize the software without limitations. This is the case with Google Earth Engine where researchers process enormous amounts of environmental data. Their data analytics algorithms can be reused by Google without restrictions since the users have to accept the company's general terms and conditions (GTC).

**5. End-user dependence on proprietary products**

User-friendly applications and platforms do not require much technical knowhow on the side of the user hiding the real complexity of data processing and analytics. When students e.g. start using Google Earth Engine they do not have to develop their own skills in data handling and transformation since the tool does not require much background knowledge of geographic information systems (GIS). Once acquainted with the comfortable online tools the users would have to start all over again if they wanted to learn how to use other platforms and tools. Thus, they are tied to a specific provider by the waste of time that switching to a competitor would entail.

**6. Technical dependence on proprietary services**

Today IT companies provide their services through web applications for end-users or application programming interfaces (API) for programmers. By integrating such APIs it is convenient to access the processed data (e.g. maps) and use it in own applications. However, this practice increases the technical dependence on such proprietary services thus raising the switching costs.

**7. Low or no-cost tools and computing power of cloud platforms**

Economies of scale allow commercial corporations to provide enormous infrastructure and data processing capabilities at very low or even no cost. This temptation of apparently free services and tools attracts governments and researchers who are able to save cost in the short run. However, it has happened in the past that private companies started with free digital offerings and then, once the diffusion was successful, switched to charging fees for the proprietary services and tools. Therefore, relying on low or no-cost offerings is a risk since their providers may change to a commercial model easily.

**8. Unique algorithmic capabilities**

With their enormous IT infrastructure (mostly within hyperscale data centers), their massive data repositories, and their skilled workforce big tech corporations and also many other IT companies are able to offer unique data processing and machine learning capabilities. Competitive research in many data-driven fields is not possible anymore without the digital services by cloud providers and IT companies. Especially in earth observation data processing there are few alternatives to proprietary platforms like Google Earth Engine, ESRI or PlanetLabs.

**C) Risks at the level of information aggregation into knowledge**

When data is processed into information, this eventually results in knowledge that has impact on decision-making, government policies, contract negotiations etc. The information is used in emergency situations, for combatting climate change, for peace missions, to increase public security etc. Therefore, it is important that the raw data is correct and that data processing is error-free because incorrect information can lead to disastrous decisions.

**9. Few big tech companies control most of the digital space**

Economies of scale as well as mergers and acquisitions have led to an oligopolistic environment where a few big tech corporations control most of the global IT market share in cloud computing and data processing. Through their market power and skilled workforce these IT corporations are able to advance new technologies rapidly. For example, currently there is a race to set standards in the 3D mapping sector. At the moment no crowd-sourced non-profit initiative like OpenStreetMap is competing. So, probably, the incumbent tech companies will soon define the new 3D cartographic standards. As a result, governments will remain dependent on a few large corporations that are continuously expanding their market dominance by extending their services and buying new startups in the data science and artificial intelligence area. Consequently, a few large IT companies remain with little alternatives besides them.

**10. Lack of transparency of the business models**

As described above, many of today's attractive digital services of IT companies are available for free. However, by definition the private sector is always based on a business model. For example, there is the established 'freemium' model in which companies offer some basic services for free. For users requiring higher performance, larger storage space, more advanced feature etc. a paid premium service is made available. In the case of environmental data processing some, however, some providers like Google with its Earth Engine do not inform about their pricing but offer their services to governments and researchers at apparently no cost. This lack of transparency about the business model raises the question of the real commercial benefits for the providers. And it increases the uncertainty among professional users about what happens to their digital assets that they operate on such platforms. For example in the case of Earth Engine, users are forced to accept general terms and conditions (GTC) that allow Google to use their data and programming scripts for any purpose.

**11. Knowhow and skill gap in public institutions**

Private companies are attractive for skilled data scientists and programmers because of pay level and career path. Therefore, the asymmetry between IT companies and governments is growing since talented people are often lured into choosing the private sector. Especially in developing and emerging countries, working for IT companies is advantageous in terms of salary. Often it is also difficult to advance a technical career in government agencies because, e.g., the organizational structures are more suitable for to administrative staff with a legal background. This results in a knowhow and skill gap in public institutions regarding data science and software development.

**12. Little awareness regarding data governance and algorithmic power**

Often, the immediate technical challenge of managing IT issues or solving data science tasks obscures the long-term effects involved with vendor lock-in. This frequent phenomenon in the IT sector describes the dependence of customers on the products or services of a certain vendor. Changing to the platform of another vendor would involve high switching costs in terms of time and money to migrate the data, change API implementations, and learn how to use a new platform. Also, cheap cloud services or apparently free tools tempt government officials and researchers to use these offerings without asking many questions. Thus, few people are aware of or care for the risks involved when private companies control data and computing infrastructure.

## 4.3 Consequences

In view of these risks, the following consequences become apparent:

**1. Monopolistic behavior**

Knowing that the users depend on their services, IT companies have much power to tighten their commercial conditions or reduce feature offerings without much risk of losing customers. Also, the lock-in effect of proprietary tools due to the lack of interoperable data formats or because of user habits allows providers to enforce rules that benefit their business perspective. Therefore, companies frequently show monopolistic behavior.

**2. Weak negotiation power**

As a result of dependence on proprietary products and in view of high switching costs, authorities and scientists using proprietary data processing services have a weak negotiation position. This is caused not only by their technical dependence (vendor lock-in) but sometimes also because of the lack of technical knowledge. This leads to ineffective regulations and loss of data sovereignty.

**3. Financial uncertainty**

When introducing new data processing tools or services, companies frequently offer them at very low or even no cost. Therefore, governments and researchers start using these services with a low budget. But once there is significant lock-in, the vendors may all of a sudden introduce fees or increase prices. This results in financial uncertainty because prices can change any time.

**4. Loss of data access**

The general terms and conditions (GTC) of cloud providers typically exclude any liability and warranty. Thus, governments, researchers and every other user may suddenly lose access to data and services in the cloud if there are no service level agreements (SLA) or similar contracts. Also, geopolitical conflicts e.g. between the United States and China or Russia may affect the availability of online data and tools like e.g. GPS.

**5. No guarantee of long-term availability**

Additionally, there is the high risk of historical data getting lost because service providers are not obliged to store it or extract it at the end of their service provision. Consequently, there is no guarantee of long-term availability of the data and their analysis. As long as no contract exists, the vendors may shut down or change their service offering anytime.

## 4.4 Recommendations

The following actions address the risks and may prevent the negative consequences:

**1. Increase awareness towards data sovereignty and digital sustainability**

Governments should support activities that increase the awareness towards data sovereignty and digital sustainability including environmental data e.g. within public procurement of new IT systems. The risks and consequences of depending on IT infrastructure and data from the private sector should be known to everyone, especially those in a leadership position and to all IT professionals and procurement officers in government.

### 2. Use, invest into and release digital public goods

Digital public goods such as open standards, open source software, crowd-sourced platforms such as Wikipedia or OpenStreetMap, open data, open machine learning models and other freely available digital assets are accessible for everyone without any limitations. Therefore, governments should benefit from these digital commons thus lowering the dependence on IT corporations. On the other hand, digital public goods are often produced by big tech corporations and private IT companies. So, they employ the skilled and experienced software developers and data scientists with knowledge about the tools and technologies. Thus, governments should not only use these freely available digital assets but also invest into and release new interoperability standards, open source software, data models and other digital assets in order to gain experience and control of these digital assets and to shape their further development.

### 3. Build and maintain publicly owned IT infrastructure

Besides virtual assets like software and data there is also the need for great computing power in order to succeed in the digital space. Big tech corporations have built enormous hyperscale data centers offering public cloud computing services for numerous tasks. For governments it makes sense to benefit from this infrastructure if there is no vendor lock-in. In order not to lose strategic control and becoming fully dependent on such cloud services, governments should invest into building and maintaining their own IT infrastructure and data processing capabilities. Long-term funding of such public IT infrastructure might be necessary since often there is no business case available. Also, publicly owned multi-state infrastructure is recommended in order to pool resources and coordinate between countries. Such multilateral initiatives allow the reduction of cost of IT spending and benefit from economies of scale.

### 4. Improve user experience and increase familiarity with public digital platforms

Besides investing into engineering, big tech corporations also spend much effort in improving user experience, knowhow-transfer and advertisement of their digital products and services. This leads to user-friendly and well-known solutions, which are highly attractive for end users. On the other hand, open source products or crowd-sourced platforms often lack usability, training and promotional activities. Therefore, governments should support the improvement of user experience and also fund training and communication activities of such non-profit platforms and communities.

### 5. Increase employer attractiveness for technical talents

Working in the private IT sector is attractive for technical talents because of high salaries, comfortable conditions (home office possibilities and still modern office space etc.), and career paths based on qualifications and performance. Often, governments are less attractive for young and ambitious people with a high technical skill level. Therefore, governments should increase their employer attractiveness e.g. by building innovative and technically challenging IT teams or change traditional structures and hierarchies to offer desirable digital career opportunities.

### 6. Remove legal barriers and increase incentives for government agencies

Several government agencies already provide and participate in public digital platforms and tools such as open source software. Nonetheless, the absence of explicit laws or the lack of specific incentives discourages the majority of government entities from using and contributing to public digital platforms. Therefore, legal barriers should be removed and smart incentives such as benchmarks of contributing agencies or innovation funds for researchers and entrepreneurs should be introduced to increase the adoption of and participation in public digital platforms. This would foster good practices and recognize positive behavior leading to a higher level of digital sustainability of public digital infrastructure.

### 7. Initiate and join organizations that support digital sustainability

Public digital platforms (like e.g. Wikipedia or OpenStreetMap) are often managed by non-profit organizations that are responsible for financial and legal issues, operate the technical infrastructure and coordinate communication and training activities. With these activities they enable the long-term availability and ongoing improvements of the public digital platforms. Therefore, governments should support existing organizations and if necessary initiate new institutions in order to ensure clear responsibilities (e.g. provide service level agreements) and funding of the digital services and products.

**8. Support entrepreneurs and researchers to create public digital goods**

There is a lack of incentives for private companies to invest into public digital goods: Their competitors may free-ride on the openly available digital assets thus reducing the benefits for the investors significantly. Also in academia there is little incentive to release research software and data and to invest into long-term availability of digital platforms since such activities do not help much for academic careers. Therefore, government should support entrepreneurs and researchers who create public digital goods. Private companies and academics could, e.g., receive additional funding when creating open data, open source software or building and maintaining digital public platforms. Business models and research activities in line with digital sustainability should be rewarded, for example, with public funds (subsidies), reduced taxes or preferred treatment in public procurement.

# 5 Conclusions

Examining the risks and consequences related to data governance and computing infrastructure used for environmental data has revealed a blind spot among governments and researchers: While the public sector benefits from innovative tools and user-friendly platforms by big tech corporations and other ICT companies, using cloud offerings usually reduces data sovereignty and increases vendor lock-in. This is no coincidence, but actually the core of the business model of the ICT industry. They bring new products and services to the market and in order to create an impetus to also use these new opportunities. The intended result is a narrative of normalcy for the use of digital solutions that make users dependent and give companies control over their data and tools.

Academics are pointing out for a long time already that the digital space is strongly controlled by the private sector. To increase the awareness of this situation, researchers Nick Couldry and Ulises Mejias have developed the concept of data colonialism. It compares today's appropriation of digital assets with the conquest of the Americas and other countries by colonial powers. In this way their concept emphasizes the need for digital alternatives that increase the autonomy of the people and governments and empower them to regain control of their data.

This report provides different insights into the daily work of experts in environmental data and digitalization issues. In the interviews, the practitioners confirm the worries of the academics that the digital space is indeed in danger of being colonized by the tech industry. However, the interviewed experts also point out ways to increase data sovereignty and digital sustainability. The report therefore concludes with a list of recommendations on how to address these problems of the digital transformation.

In essence, the focus on using and creating digital public goods and the capability to build and maintain publicly owned IT infrastructure is also the foundation of the concept of digital sustainability. Increased public investments allow improving user experience and familiarity with such open source and open data platforms. Also, it is necessary to attract talented engineers and product managers to serve in the non-profit sector towards the benefit of the society. Existing initiatives and organizations should be supported by governmental and research funding. And social entrepreneurs and academics could be incentivized to create public digital goods. This makes it possible to increase digital sustainability and data sovereignty of the digital space and thus improve the prospects for future generations.

Future investigations could focus on three different issues: First, the current list of conversations should be extended with additional experts recommended by the interviewed persons. For instance, the interviewees suggested exploring the eu-LISA and the Swiss Data Cube platforms, two technological initiatives driven by public agencies. It would be interesting to develop case studies of these initiatives and elaborate success factors, problems, and possible solutions. Second, the scope of environmental data and its use should be analyzed in order to describe additional examples of data-driven activities. In this way, the concept of data colonialism could be further explored and enhanced with relevant examples. It would also allow extending the analysis towards the environmental impact of digitalization. The new initiative "Coalition for Digital

Environmental Sustainability" (CODES) by the United Nations Environment Program (UNEP) and other organizations focuses on such issues (UNEP et al., 2021). And third, the recommended solutions in this report should be assessed towards their practical feasibility and challenged with possible new risks and involved costs. In addition, such an assessment would allow finding additional measures towards a more sustainable and sovereign management of environmental data. As the digital world and the amount of data is constantly evolving, it is of great importance to further identify opportunities and risks and to implement new standards to regulate its usage and access.

# Appendix A: Semi-structured interview guideline

**A) Questions on data management (individual perspective)**

1. Who collects the data you use? Who controls it?
2. What cloud platforms do you use to manage and store the data you use?
3. What software tools are relevant for your work?
4. Is there an asymmetry between governments and corporations collecting and managing data?
5. Did you have certain problems in the past getting access to data relevant to your professional activities?
6. If yes, was the data not available at all, was it in control by private companies or in what other way did you have issues?
7. How are data and platforms interoperable today? What are issues involved in data transfer?
8. How did you solve these problems with data?

**B) Questions on governmental data access and data use (global perspective)**

1. How do governments use environmental data? Who owns or controls this data?
2. In what way are governments dependent on data by corporations? (e.g. geopolitical effects)
3. How do digital skills of employees within governments vs. companies to this issue?
4. What are the business models of corporations producing and providing data?
5. What problems do you observe involved with control of data by certain countries or by companies?
6. How do governments or corporations misuse data today? What is the potential of data misuse?
7. What are the different data privacy strategies of national governments?
8. In particular, what is the role of the US and Chinese government regarding their big tech corporations?

**C) Questions on digital sustainability, digital sovereignty and other potential solutions**

1. What organizations and initiatives do you know regarding digital sustainability, sovereignty etc.?
2. Have you heard of concepts like "data colonialism", "data sovereignty" or "digital sustainability"?
3. Do you think these concepts are relevant today or in the future? If yes, why? If no, why not?
4. How could your organization or how can governments benefit from data sovereignty or digital sustainability?
5. What could your organization or what could governments do to increase data sovereignty?
6. Do you use or produce digital public goods? If yes, how? If no, why not?
7. What other solutions to address the above discussed problems do you see?
8. How could open resp. crowd-sourced data platforms such as OpenStreetMap improve this situation? Should they become mandatory for governmental ties?
9. What potential risks and challenges derive from open data platforms?

# Appendix B: Interview Summaries

## Interview 1: David Jensen, UN Environment Program

David Jensen is the Head of the Environmental Peacebuilding Program at UN Environment (UNEP). Since 2009, David has been a leader in a global effort to establish a new multidisciplinary field of environmental peacebuilding. This field aims to promote environmental and natural resource management to prevent, mitigate, resolve, and recover from conflict. It also seeks to use shared dependence on natural resources and ecosystems as a platform for cooperation and confidence building among communities and countries.

### Power of the big tech industry

There is a shift in terms of environmental data and analysis from the public to the private sector. This shift happens in terms of who is aggregating the data, who is processing it, who is analyzing, and who is publishing it. Corporations are controlling more and more the infrastructure, the derived analytics and also more of the narrative. This trend is going to get worse and the public sector is becoming increasingly dependent on the private sector for global-level environmental monitoring and analysis. And there is an asymmetry regarding the knowledge and technical power of corporations, which is growing, and we don't always know how to negotiate with them because we don't know what they might do with some of the derived data products they are working on and how they might plan to monetize them. Big tech companies are offering digital services to all these public actors and amassing huge piles of environmental data and we don't really know what they are going to do with that behind the scenes because they aren't always transparent about their intentions. We are giving them a massive quantity of intelligence in exchange for very little and there are no sustainable business models in place to fund the collection and management of these digital public goods. They are trying to attract a lot of public actors and a lot of non-governmental actors into their ecosystems on the basis of "oh this is the next big thing" and "we are going to help you process your data", but ultimately, they are after aggregating as much data as they can in order to eventually mine and sell derivative analysis and insights. More transparency about their intent is needed together with some form of revenue sharing and business model so that the costs of the digital public goods can be covered.

### Asymmetries in competences

The private sector can often attract the best AI scientists, the best data scientists, the best analysts due to the high salaries. The capacity of the public sector is at risk because a lot of talented people in the public space are moving private. This is happening in particular in many African countries. There is a growing asymmetry in terms of expertise between private sector and public organizations. When public actors are negotiating partnership agreements or data access or anything, the companies just know far more than public actors about potential use cases and applications. They know how they can deploy these data analytics in ways public actors can't even imagine. This asymmetry is used against public actors. It has always been the case that the private sector is faster than the public sector, but what has changed is the scale and the speed. They have global reach now and are able to manipulate human activities to a level that we are not aware of. The lack of transparency in the underlying business models of the digital sector is a massive problem.

### Risks of digitalization

One of the current challenges public actors face is the commercialization of digital public goods, in particular public good data. What data sets are the companies combining and what are they extracting? How are they commercializing it? Then there are the human resources: The public sector is being emptied out from our capacities, because the private sector is pulling in. A lot of the AI and data science people are being attracted into the private sector and it is really hard to compete with the salaries and benefits the private sector can offer. Then the geopolitical situation: Having the whole world storing the data in clouds mainly in one jurisdiction in the US is risky. The global data is being subjected to laws dictated by a single country. From a legal perspective, a big tech company might not be able to mine its own databases, but a government certainly could if following national law. They could dip into the whole tech infrastructure of big tech companies and extract intelligence if they wanted to and that is a pretty big risk. At the moment, the US political and economic power is largely a consequence of their technology sector and the lack of governance of it. Another risk is cyber security: We no longer understand, what kind of backdoors are put into our digital technologies. We could have more cloud services in our own country as, e.g., in Switzerland with many cloud storages, but the IT infrastructure and components which are needed for this are coming from other countries and we have no idea what is in their technology at the chip level. Also, there will be upcoming economic asymmetries in terms of forecasting: Countries or companies that are developing and using artificial intelligence first will be able to predict stock markets and other market developments and thus have advantage over the others.

**Data colonialism and data sovereignty**

The UNEP could be accused of data colonialism in terms of our global surface water explorer, where we are monitoring the extent of freshwater around the world using satellite data, cloud computing and AI, without getting the consent of the countries in advance. We seek their support in terms of publishing the resulting analytics and give countries the opportunity to opt out of the analysis. But given the fact that UNEP doesn't control the satellites, countries are not given a chance to opt of the data collection phase. To address issues related to data sovereignty and governance, we need to broker agreements between private sector and governments to agree on what are the core high-value data sets needed to monitor global progress against the indicators of the Sustainable Development Goals (SDGs) and the Multi-lateral Environmental Agreements (MEAs) that should be digital public goods. And then agree on a financing mechanism and business models for these digital public goods and ensure they are published in an open and inter-operable format.

While some might think that everything about the biophysical description of the environment and natural resources on earth should be public, we also need to think about potential risks. One might not want to make the location of endangered wildlife populations public, because that could put them at risk. The same holds for uranium deposits, because there is a certain security risk. We need to determine which global data sets should be public and which ones are needed by public and private sector actors to accelerate action in achieving the SDGs and MEAs? Forest cover, land cover, water cover should be public and there needs to be a way to finance that. This applies also to the distribution and trends of natural resources.

## Interview 2: Matthias Leese, Center for Security Studies at ETH Zürich

Matthias Leese is researching how the police work with digital data for Predictive Policing and therefore he has conducted many interviews on this subject with members of the police. He found that the police mainly use commercially available products even though there has been a tendency in the last few years to develop own products to ensure to control the processes and to control leaks. The only commercially available product that is usable out-of-the-box is from German manufacturer IfmPT (Institut für musterbasierte Prognosetechnologie/Institute for pattern-based prognosis technology). They aggregate historic data on crime and derive data on which neighborhoods are vulnerable to which type of crime. When police decide to use this product they are obligated to hand over data from the last 5-10 years for a retrospective analysis of vulnerability which is then employed as basis for pattern analysis.

**Risks of data abuse**

There are two risks: Stigmatization of neighborhoods as "criminal hotspots" even though this is objectively not true. There are also aspects of data security. When there is data on a private server there is a risk that non-authorized people can gain access. In Germany and Switzerland the police do their own analyses and enrich them with data from third-party providers. This is in contrast to the USA where there are online platforms like HunchLab where data can be analyzed and the databases can be synchronized locally and nationally. Platforms like these are unlikely to come into use in Germany and Switzerland due to differences in culture and also due to the General Data Protection Regulation (GDPR). Data security and data sovereignty are regarded as highly important.

**How Switzerland is intertwined with the EU**

Switzerland is part of the Schengen agreement and is therefore also part of the Schengen Information System, which is the largest database in the EU to share police data. There are alerts like warrants, stolen objects, missing persons, missing objects, forged papers. All the Schengen members feed into this database. It is an inter-state project and it is managed by eu-LISA (European Agency for the operational management of large-scale IT Systems in the area of freedom, security and justice).

### Cooperation with private parties

One big advantage of cooperating with the private sector is cost-benefit analysis and expertise: There is good software on the market with a good user experience design, good software-support with regular updates and developments. Additionally, there is someone responsible who can be held accountable should something not work as desired. The big disadvantage is that data sovereignty is sacrificed and that standard software has to be adapted. For this, expertise has to be bought which raises the question of competitive salaries and competitive career-opportunities compared to the private sector.

### Outlook

Open Data, Open Government, and Open Research Data are very important concepts. But there is the issue of where one has to draw the line, which data should be made public and which not. And who is, eventually, responsible for the data?

## Interview 3: Anna Brach, Geneva Centre for Security Policy

Anna Brach researches secondary sources in order to provide the best information for her students who come from different governments all over the world. As she teaches about environmental security and climate change her data comes, for example, from the Intergovernmental Panel on Climate Change or from the World Resource Institute.

### Data on water

The official sources of water data are the UN Environment Program and the World Resource Institute. Data on water is important in the context of climate change. It is about the quantity and quality of water, which is influenced by climate change and can contribute to geopolitical issues and conflicts. Water always has been a strategic resource but it is becoming increasingly important with the on-going climate change, especially data on water availability for particular countries when it comes to shared water. But it is also important for strategic decision-making. The importance of real-time environmental data is increasing, especially during conflicts.

### Open data

There are some countries that are reluctant to share their data. More democratic countries are more willing to share their data, and especially those who believe in multilateralism. The biggest problem is not that data is not available, though, but it is the way certain countries communicate. For example when they claim to aim at becoming carbon-neutral but are building coal-fired power plants at the same time.

## Interview 4: Fritz Brugger, NADEL at ETH Zürich

Fritz Brugger of the Center for Development and Cooperation (NADEL) at ETH Zürich is specialized in the developmental effects of natural resource extraction and on the role of mining companies. He gathers data in the form of household surveys stored on own servers. He and his team also collect data, process it and then offer it for local governance processes. With this they contribute to the political dialogue in developing countries.

### Environmental data

Environmental data means all data that is concerned with the environment. This includes data about water quality, satellite data, weather data, and measurements of flow rates. An interesting example are conflicts at river basins that extend over several countries. A common element in transboundary water conflicts is the unwillingness of governments to share data; this is a political problem, often fueled by mistrust towards other parties, not a technical problem in the first place. One important step in conflict mediation is to foster the willingness of parties for a minimal data sharing on a technical level.

### Outsourcing services

When working with data, it is important to recognize the risks along what we call the "data value chain", i.e. collection, storage, processing which transforms raw data into information that informs decisions. The risk to be considered include questions of sustainability (financial, knowledge, privacy, security). The data collection typically happens locally. Then data are often transferred out of the country or to private operators for processing or storage. This appears advantageous since you do not have to provide the infrastructure or the knowledge necessary. But you become dependent on the external service provider and the question is whether you have the skills and abilities to adequately assess the quality of this service provider. If there is not sufficient expertise on the side of the data owner, risks and dependency increase. Governments or development agencies also may have concerns over data-sovereignty when storing certain data out of the country .

**Data interoperability**

Data interoperability is a common technical hurdle. With this problem, technology is usually only 20 percent of the solution. 80 percent of the solution is agreement on the political level, who shall have access to which data what always also means which means interests, influence, control, hidden agendas, priorities, 'who tells whom what to do.' Ideas like Open Government Data are all nice and good, but the reality is simply different. The key question is 'who gets to see and use the data?" and that's where power and control come in.

**Digital skill gaps**

When you work with IT tools, you automatically prioritize people who have access. These are the higher educated urban men. This introduces a bias and the risk that existing inequalities are not eliminated but will be exacerbated. But the skill gap continues between cultures. Ghana and Kenya, after all, have their own developer scenes that are slowly forming. But at the same time, there are clearly Western financiers in the background.

There are also digital skill gaps between people working in governments and those in private companies. Generally, there are hardly any job opportunities for developers in the public sector; so they go where they can do something and get paid well for it.

**Data colonialism**

Data colonialism is a reality and not just a story between North and South, but also regarding platforms and networks. The World Development Report 2016 - Digital Dividends summarizes this very well. A key element here is data privacy laws that differ vastly, also between e.g. the EU and the US. ETH Zürich, e.g., has a policy that prohibits using services that are hosted in the US. For example they do not use Dropbox but Polybox, open source software hosted on ETH Zürich servers.

## Interview 5: Stefan Keller, Eastern Switzerland University of Applied Sciences

Stefan Keller is an expert on data engineering and spatial data analytics. He is generally interested in making OpenStreetMap data useable for authorities and, vice-versa, to make data from authorities useable for OpenStreetMap. He operates a geo-converter and a downloader for the conversion of OpenStreetMap data to GIS-data as a service. To him environmental data are spatial data that not only concern topography but also, e.g., weather and climate.

**Risks of data abuse**

The main risks are when there are third-party data at the beginning of the planning-decision-cycle that a private company can change or deny access to. Such a situation leads to problems in the whole process. Control and plausibility of data are very important, and also that the raw data is transparent at every moment. GPS/GNSS are US-American monopolies and there are examples where the data was or is interfered with (e.g., the Afghan war where GPS was less precise on purpose; and in China GPS is shifted a few hundred meters to ensure Chinese mapping sovereignty).

**Private firms and geo data**

In mapping technology there are the US-American TomTom and the European TeleAtlas. These companies and Google have started to build their own maps. Now there is also the 3D-market with big players like Niantic, Apple, Facebook, and Amazon. Then there is the industry for satellite sensors, which is in a crisis at the moment because they cannot generate enough value with their data. Therefore the satellite industry has started a useability initiative by making their data useable by providing analysis-ready satellite data.

NASA recently introduced a new project where they aggregated their satellite programs with a unified 30m-grid with programming-useability-friendly repository and web-interface. There is also Sentinel-data from ESA where everything is in the same format, which is an absolute milestone.

In the 2D-sector authorities use OpenStreetMap more and more. It is already standard for less commercially interesting areas like humanitarian help. In the 3D-sector there is no initiative yet around. Without an open crowd-sourcing platform one of the big players is going to establish a solution. This would mean dependence on their services.

**Data colonialism**

It is crucial that people become aware that they are the owners of their data. And authorities should become more sensitive when using commercial services and should e.g. not use Google Maps. To avoid dependencies, it is important that data is freely accessible in its full detail like it's the case with OpenStreetMap. It is also important that authorities have enough technical competence in order deliver (geospatial) base services, so they do not have to delegate everything to third-party companies.

## Interview 6: Thomas Schneider, Swiss Federal Office of Communications

The Swiss Federal Office of Communications (OFCOM) represents Switzerland in key areas of international internet governance and digital cooperation. In the past years, new topics such as artificial intelligence or digital self-determination were added and OFCOM is also representing Switzerland in these fields. It is also responsible for the monitoring of the implementation of the strategic guidelines on this of the Swiss Confederation. In close cooperation with the Federal Department of Foreign Affairs (FDFA), OFCOM has been working on the development of the global internet and digital governance architecture, with the involvement of all stakeholders.

**Data governance**

At the 2005 UN World Summit on the information society, internet governance was defined as the totality of all norms like laws, international agreements, technical standards, economic regulations, social and unwritten norms shaping the use and further development of the internet. A similar approach can be used with describing data governance. For example, most digital applications and social media are developed by young, white and male people from western countries. They are not representative of their societies and often ignore the situation in developing countries. Their mindset can cause certain services to have undesirable negative consequences for population groups that were not included in the development process. Another issue is how to deal with social media when it comes to incitement of violence. Here, measures taken by online platforms themselves as well as regulators have to respect freedom of expression and protection of personal safety at the same time and find an appropriate balance in case of rights conflicting and also between different cultural traditions.

**Risks**

It is essential for a society to keep its digital sovereignty and for individuals their digital self-determination. If a country does not control the hardware, software, or technology itself, its national sovereignty is at risk. It is essential to know how a technology works and who can intercept or manipulate information. A society does not have to produce everything itself, but it needs to understand it. And we need to be connected with others, so that, together, we can keep control over the key functions of our societies. Another risk often cited is the danger of fragmentation of our societies, when everyone lives in their own bubble of like-minded peers. An open public debate culture and thus the ability to find compromises between diverging views can be lost or at least become very difficult. An example of this is the polarization that we currently witness in countries like the US, where we have conflicting narratives that become "alternative truths" that make it difficult for the political system to function. Another risk is actually coming with seizing opportunities: it makes sense to use data to improve our lives as well as our environmental situation, e.g. for making the fight climate change and other challenges more efficient. But the more we rely on data in the functioning of our lives, societies and economies, the higher the risk that we dependent on global players who are not necessarily committed to the benefit of a national society or whom you cannot prescribe the rules of the game on national level. This is relevant for data in the health sector, judicial system, social security, mobility and others. As a consequence, synergies and economies of scale lead to monopolistic tendencies that give a lot of power and influence to a small group of big global actors. .

### Dependencies

Apart from trying to do everything oneself, key pillars to reduce dependence are, first, avoid monopolies and foster choice through competition: If competition prevails, one can choose companies for one's services that correspond to one's own values and expectations in dealing with the data. On the individual level, consumers need to be aware that their decisions also impact the market situation and thus their future choice. A second pillar is to ensure democratic legitimacy over the use of data and "natural monopolies". Monopolies usually lead to abusing power. Democratic governance and control over a monopoly usually reduces the risk of abuse. Creating the right incentives is also relevant for Open Government Data (OGD): When e.g., the SBB and Swisscom make their data available, access to their data should be organized in a way that not only global players but also small and local actors are actually able to use it, otherwise dependence might not be reduced but increased. OGD is not good just because it is open. It only unfolds positive effects when the incentives are set in such a way that it does not simply support the growth of the big players but that new market players also have a chance.

### Digital skills

Big players often buy the biggest talents right out of the market – in all countries. They often pick them up directly from universities. This problem is not new. It also exists in other areas, but it is more accentuated in the IT realm. This is a challenge to smaller private actors as well as for the public administration. In order to compensate the scarcity of means, public authorities have to connect and cooperate horizontally among themselves on national and international levels. And they have to cooperate and connect with the industry, academia, civil society and the technical community and foster the development of a multistakeholder dialogue culture that allows for a constructive exchange and a definition and respect of the respective roles of all stakeholders.

### Data colonialism and data sovereignty

If powerful private actors or states gain control over other societies and are able to impose rules on them, this may be called data colonialism. To prevent monopolies and dependencies we need a discussion and concepts on what data sovereignty should be (and who the "sovereign" in this regard should be) and how to achieve this. From a democratic point of view, individual digital self-determination must be at the core of every concept of digital sovereignty. And it is crucial that smaller players and societies work together to jointly defend their sovereignty and self-determination. This is also important for reasons of equality and sustainability.

## Interview 7: Andreas Heinimann, University of Bern

Andreas Heinimann worked on democratizing access to data on land issues and also improve data quality big project sis in Myanmar and Laos over the last decade. If land issues are not transparent, decisions are a question of power: The stronger party obtains land and resources. They have worked together with 24 departments as well as civil society in Myanmar and improve the respective data and especially their cross sectorial coherence. They have documented changes in the environment using satellite data. Big environmental data to him are mainly Earth Observation (EO) based data: such as vegetation, data on water, water availability, pollution, climate etc.

### Risks of data abuse

One big issue is transparency. In the global south borders are not always clear and due to this there are regional tensions. When one makes data transparent, there problems may arise because inconsistencies in datasets become visible. The data custodian, in this case Myanmar, therefore does not want to share the data. If non-perfect data becomes public a department can lose legitimacy. Before the recent political crisis in Myanmar, Heinimann and his team have worked together with the departments to improve their data so the main obstacle to making them public can be avoided. Data always hold the danger of abuse and misinterpretation.

**Dependence on Google**

When working with spatial data there is the risk of becoming dependent on the Google Earth Engine. This is a free to use platform availing immense amounts of public domain EO data offering cloud computing ability. This has changed the way EO data can be analyzed and what insights can be produced. For example, a high resolution global deforestation analysis became possible where the analysis of 700'000 satellite images took 100 hours instead of 100'000 hours on a workstation. This service is offered free. The problem is that the service is not transparent. For example, there is no information on how many pixels can be processed. When using Google Earth Engine the user actually works as a developer for Google, as the developed scripts are available to Google. Even more importantly, there is no transparent or public stewardship of the platform, and there is no guarantee of continuity. Google can basically decide to just stop this service.. This is why it is important that there are alternative infrastructures on the level of state-associations (e.g., EU).

**Data sovereignty and data colonialism**

Data Sovereignty and making explicit of data custodianship is important. Worse than no access is access with no transparency. People generating data often do not think about who actually owns the data and whether they violate data sovereignty. For example, if data that local communities have collected/produced (e.g. through participatory mapping) is made available online without making 100% sure that the communities full understand this and gave their explicit consent. Concerning data colonialism there is a new digital divide: Those who provide data have to be compensated for it.

## Interview 8: Nick Couldry, London School of Economics and Political Science, and Ulises Mejias, State University of New York

Nick Couldry and Ulises Mejias are the co-authors of the book "The Costs of Connections – How Data Is Colonizing Human Life and Appropriating it for Capitalism".

**Data colonialism**

Data Colonialism is not a metaphor, it describes specifically this development of an emerging social order with specific characteristics and with historical continuities. A key part of the argument is the difference between colonialism and coloniality. Even though colonialism might be over, the legacy of colonialism continues. So when we look at certain legacies of power structure, gender relations and inequality between nations that legacy of course continues. The core of colonialism was not the grabbing of land as such, it was the idea that the west has the knowledge and superior rationality, and thus rationalized the grabbing of the land. It was the attempt to dominate with one version of knowledge, when in fact the world has many forms of knowledges, many perspectives. And this new type of knowledge, which dominates, is data.

In the historical colonialism, nature was defined as 'cheap', it is just there for the taking, it does not belong to anybody. The next step was to come up with the concept of 'cheap labor'. Because to extract all those resources from nature, a large supply of cheap human labor is needed achieved through racism. So nowadays we have 'cheap data', which has many of the same characteristics: it is abundant, it is everywhere and free to take for everybody. We all generate it, but only some have the technology to process and make use of it.

Up until the eighties governments possessed most of the information about their territories. Now corporations have millions of times more information than the government. So governments now rely on information from corporations and that obviously has implications on the power relations between them.

**Digital skills**

Digital skills have to be more evenly shared. The standard curriculum has to be changed because the way we train our engineers and our scientists makes a huge difference. It is not just that more people should learn to code, it is also important that the coders need to learn other skills and widen their world view. Coding algorithms is not just a way to get things done, it is also part of a process of reproducing the world. This means it has to be done in a sustainable way, which includes the environmental costs, but also the costs of the social environment, i.e. our relations to each other. A lot of computer scientists do not understand that they are actually reengineering the world with the project they are working on. And sometimes that is not a healthy or sustainable project.

### Geographic data

In the current climate crisis, we need as much environmental data as possible in order to monitor the crisis. At the same time, many corporations such as Microsoft and IBM are tempted to profit from the situation. When we look at some of the initiatives of Facebook we see a very intentional use of developing tools to extract data from the global south and to use that data for the benefit of the corporation. Google Earth Engine is part of the problem – in the South they simply do not have the resources to build their own systems.

## Interview 9: Fatine Ezbakhe, University of Geneva and Geneva Water Hub

Fatine Ezbakhe has an engineering background and works with data and tools for sustainable development, particularly for water. She has been at the Geneva Water Hub since 2019 and her current research looks at the political dimension water resources and the role of data in shaping (and being shaped) by politics. She is interested in the link between data and information. In her PhD, she focused on the issue of statistical uncertainty and how to deal with it in decision analysis. One of the main sources of information in the water, sanitation and hygiene (WASH) sector are household surveys, which ask households about their access to water and sanitation and the quality of the services. This information is then employed to identify underserved areas and prioritize them in future investments. That is why data is needed for decision-making and better management of water resources and services.

### Environmental data

Data is needed to measure, model, or monitor various elements of the environment. This includes water-related data (e.g., precipitation, surface runoff, quality parameters) to understand and model the water cycle, and also socio-economic data (e.g., population trends, land use, societal wellbeing) because water resources serve different socio-economic sectors. When it comes to river basins, the governments and various institutions such as river basin organizations (RBOs) are responsible for managing and monitoring the rivers. They collect and analyze the data, and sometimes they make it available to the public. The private sector is also involved, for example, service providers or hydropower companies. The institutions and companies Fatine Ezbakhe is familiar with generally share yearly reports, not the raw data. Furthermore, data itself becomes an important strategic resource. One of the key issues is the exchange of data. The lack of data leads to a lack of knowledge about what is happening along the river, which adds a layer of difficulties in planning the infrastructure and its use. That is why now we tend to see the data as a way to build trust, which might lead to more cooperation over water. For this reason, at the global level, data exchange has become an important aspect of water management.

### Asymmetries

Lately, there has been a lot of talk (and initiatives) about using satellite data for water/environmental monitoring. When talking about satellite data and remote sensing, some question the asymmetries in the data collection process, because the data providers do not always match with the data owners. This happens, for instance, when the software used for water monitoring is not owned by the actors using it, which can lead to issues of restrictive licenses. Furthermore, in the case of transboundary water resources – i.e., those crossing political borders – data asymmetry can happen between upstream and downstream countries. At the local scale, there is also asymmetry. It is not always clear whether the people of the questioned households gave their consent to make their data available at global repositories. Even though the data is aggregated and the specific household is not directly visible, there is the risk of having data disaggregated down to the very local level. And then you might be able to identify the household, for example, if there is a town with only five households and all of them answered that they have no toilet. This data is not only numbers; it is data that affects their personal lives and privacy rights.

**Consent**

One risk is to forget the (informed) consent. For example, smart water meters installed in our households collect personal data such as water consumption, geospatial location and payment information. Water utilities use this data to manage the water distribution network. By giving consent and knowing where your data goes, you are becoming part of the data cycle. But with consent also comes the debate about compensation. The rationale is that, since the data provider is part of the data cycle, she or he could be compensated for the value of her or his data. This could also be in non-monetary terms, like making them a meaningful part of the cycle that involves the data that they give.